# An Improved XP Software Development Process Model

M. R. J. Qureshi and S.A. Hussain
Dept. of Computer Science,
COMSATS Institute of Information Technology Lahore Pakistan
rjamil@ciitlahore.edu.pk, asadhussain@ciitlahore.edu.pk
Ph # (92-42-5431602) Cell # (03334492203)

## Abstract
*The concept of agile process models has attained great popularity in software (SW) development community in last few years. Agile models promote fast development. Fast development has certain drawbacks, such as weak documentation and performance for medium and large development projects. Fast development also promotes use of agile process models in small-scale projects. This paper modifies and evaluates Extreme Programming (XP) process model and proposes a novel process model based on these modifications.*

## I. Introduction

Agile process models stress on agility for software development. Agility signifies responding to changes quickly and efficiently. Possible changes required in software projects are in budget, schedule, resources, technology, requirements and team. These are "reacting" changes on which agile models stress. They are called agile golden principles that are defined in agile alliance meeting conducted in 2001 [1].

The aim of agile principles is to have adaptive software development only for simple and small size software projects. There is no indication to adapt process models according to nature of the projects. Analyst has to select traditional software process models if the software is average or complex in nature, such as Spiral and Rational Unified Process (RUP). Section 2 of this paper describes related work about agile models. Section 3 proposes an improved XP process model for agile and traditional software development. Section 4 describes main features of improved XP process model

## II. Related Work

A number of case studies have been reported in last few years to support agile process models. These papers are written to support agile process models [2,3,4,5,6].

- Agile teams are supposed to monitor team performance and SW development procedures continuously to perform better and efficient.
- Results of case studies show that defect rates and team's overall productivity are improved by following agile practices.

The authors provided three case studies to provide practical proofs [2,3,4,5,6]. Completion time of two of these case studies was three months and the other one was to be completed in eight weeks. Both qualitative and quantitative techniques were used to estimate and analyze the results of these three projects. These papers were written to adapt agile projects according to agile principles. The aim is to adapt XP model for two to three months projects. These papers lacked the knowledge for adaptation of agile XP model for medium and large projects (greater than 1000 function points).

Lucas Layman, et al [7,8,9], adapted XP model for two case studies. One case study was conducted with IBM for one year and other was conducted with Sabre Airline Solutions for three months project. The authors proposed an extreme programming evaluation framework (XP-EF) to adapt XP model. The framework used feedback loop to permit the agile team and procedures to get better. Framework needs further validation through more case studies and needs to be improved specially with respect to the XP adherence metrics. The teams in both case studies were typically agile and have management support to employ the XP process model. Hence, success stories of both teams can not be taken as an example to fit for non agile, large and distributed teams and for teams that do not have management support to use XP process model.

## III. An Improved XP Process Model for Agile Development

Adaptive process model is a modified approach of XP model, which is most widely practiced model among agile models. The main phases of XP model are planning, design, coding and testing. The main phases of adaptive model are Project Planning, Analysis and Risk Management, Design & Development and Testing.

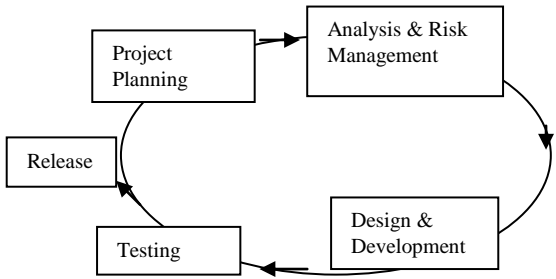

**Figure 1 An Improved XP Process Model**

The focus of the proposed process model is the implementation on medium and large scale projects which keep on evolving due to changes in customer requirements. Project specification or proposal document is prepared during the *'Project Planning'* phase by communicating to the customer. Project specification or proposal document is composed of feasibility report that is created to prepare a cost benefits analysis (CBA) sheet. Feasibility report is composed of economic, technical and operational feasibilities. Organizational feasibility is also prepared based on client request or if project demands. CBA sheet helps to estimate whether SW project is feasible for the customer or not. Project team members are also selected during planning phase. Project team size depends on the size and schedule of project.

*'Analysis & Risk Management'* phase is only started if a customer approves the proposal. Analysis phase improves quality of software through proper documentation. This is the phase in which an analyst gathers detailed requirements. Software requirement specification (SRS) document is produced at the end of analysis phase. Main contents of SRS are: summary of requirements, requirements modeling, data modeling and risk management plan. Requirements could either be structured or object oriented (OO). Entity relationship diagram (ERD) is drawn in case of structured development and object diagram is created in case of object oriented development.

*Design and development* phases of XP model are merged to incorporate agility in the proposed process model. The improved XP process model uses the prototype approach to verify the design and requirements. Software is developed incrementally as customer approves prototypes.

Test cases are prepared for each increment at the start of *Testing* phase. Each module is tested on unit basis. Integration test is then performed to check integration among modules.

System test is the next phase to validate the whole increment as one unit. Acceptance testing is the last test to verify increment from the customer. Tested increment is maintained and deployed. The proposed process model is cyclic and evolutionary till whole software is developed.

## IV. Main Features of Improved XP Model

Main features of the Improved XP process model are as follows.

- A new framework in software engineering field.
- It is multidimensional in nature. Improved XP model is equally suitable for incremental and parallel development according to nature of project.
- 'Analysis and Risk Management' phase in the new framework has many benefits e.g., initial risk management plan to cater the potential risks regarding the failure of project.
- 'Design and Development' phase helps to keep the new framework agile in nature.

The new framework provides:

- Support for component-based development.
- Support for distributed development environment.
- Support to have large team size.
- Support for subcontracting.
- Better documentation helping to software engineering team during and after development (e.g., Risk Management).

Analysis phase results in more comprehensive user requirements, modeling and documentation. SRS is the end product of analysis phase. Comprehensive analysis always gives better design which results in high quality software. Risk management plan at analysis phase is another strong feature of improved XP model as compared to XP model. There were so many projects failed because of lack of risk management when XP model was proposed initially in 2001 [10,11].

Design and development phases of XP process model have been merged in the improved XP model. The objective is to achieve agility in the proposed model. Prototype approach helps to verify the design and requirements. Merging of design and development also improves efficiency of the software.

Main purpose of the improved XP process model is to eliminate prominent limitations of agile process models and particularly of XP. The improved XP process model is proposed to support CBD, large teams, complex and safety critical project development. The addition of proper analysis phase also helps an analyst to consider the factor of reuse during modeling a system. Reusability is an important advantage of developing SW applications using CBD. CBD helps to reuse components frequently in the similar applications that result in time and cost savings [12]. Analysis phase also results in efficient process design. Documentation and quality are significantly improved because of incorporation of these changes. Removal of one limitation eliminates all since all of these limitations are interlinked.

Analyst can adapt the improved XP model according to the nature of software projects. The proposed model can be implemented for agile, medium and large projects.

## V. Conclusion

This paper supports practice of agile software development by proposing a process model which is adapted according to the requirements of the software project. The improved XP process model is better than XP because it eliminates the limitations of development of reusable components, large development teams, documentation, medium and large software projects.